\documentclass[%
 reprint,
 amsmath,amssymb,
 aps,
]{revtex4-1}
\usepackage{amsthm}
\usepackage{graphicx}
\usepackage{dcolumn}
\usepackage{bm}
\usepackage{extarrows}
\usepackage{float}
\usepackage{graphicx}
\usepackage[roman]{complexity}

\begin{document}

\preprint{APS/123-QED}

\title{Variational Quantum algorithm for Poisson equation}

\author{Hailing Liu $^{1,2}$}

\author{Yusen Wu $^{1}$}

\author{Linchun Wan $^{1}$}

\author{Shijie Pan $^{1}$}

\author{Sujuan Qin $^{1}$}
\email{qsujuan@bupt.edu.cn}

\author{Fei Gao $^{1,3}$}
\email{gaof@bupt.edu.cn}

\author{Qiaoyan Wen $^{1}$}
\email{wqy@bupt.edu.cn}

\affiliation{$^{1}$ State Key Laboratory of Networking and Switching Technology, Beijing University of Posts and Telecommunications, Beijing 100876, China}
\affiliation{$^{2}$ State Key Laboratory of Cryptology, P.O. Box 5159, Beijing, 100878, China}
\affiliation{ $^{3}$ Center for Quantum Computing, Peng Cheng Laboratory, Shenzhen 518055, China}

\date{\today}

\begin{abstract}
The Poisson equation has wide applications in many areas of science and engineering. Although there are some quantum algorithms that can efficiently solve the Poisson equation, they generally require a fault-tolerant quantum computer which is beyond the current technology. In this paper, we propose a Variational Quantum Algorithm (VQA) to solve the Poisson equation, which can be executed on Noise Intermediate-Scale Quantum (NISQ) devices. In detail, we first adopt the finite difference method to transform the Poisson equation into a linear system. Then, according to the special structure of the linear system, we find an explicit tensor product decomposition, with only $2\log n+1$ items, of its coefficient matrix under a specific set of simple operators, where $n$ is the dimension of the coefficient matrix. This implies that the proposed VQA only needs $O(\log n)$ measurements, which dramatically reduce quantum resources. Additionally, we perform quantum Bell measurements to efficiently evaluate the expectation values of simple operators. Numerical experiments demonstrate that our algorithm can effectively solve the Poisson equation.
\end{abstract}

\pacs{Valid PACS appear here}
\maketitle


\section{Introduction}

Quantum computing has been shown to be more computationally powerful over classical computing in solving certain problems, such as factoring large numbers \cite{S1994}, unstructured database
searching \cite{LKG1997}, solving equations \cite{HHL2009,WYPGWQ2018}, classification\cite{PMSL2014,BJYD2017}, linear regression \cite{NDS2012,YGW2019}, and dimensionality reduction \cite{SMP2014,IL2016,PWL2020}.

The Poisson equation has wide applications in many areas, such as quantum mechanical continuum solvation \cite{TMC2005}, and Markov chains \cite{MSP2007,AG2007}. In general, the finite difference and spectral method \cite{FGEW2008,C2007,ELC1998} are used to solve the Poisson equation. The core of these algorithms is to approximate the solution of the Poisson equation with the solution of linear systems. Since the dimension of the linear system obtained by the discrete Poisson equation is generally very large, solving the linear system is quite time consuming. In order to solve the Poisson equation efficiently, some related quantum algorithms \cite{CPP2019,CLO2020} were proposed. These quantum algorithms have shown significant speedups over their classical counterparts.

However, the advantages of quantum algorithms mentioned above usually rely on a fault-tolerant quantum computer, which may take a long time horizon to implement. Recent developments in quantum hardware have motivated advances in algorithms to run in the so-called Noisy Intermediate Scale Quantum (NISQ) devices \cite{P2018} which only support a shallow quantum circuit, a restricted number of physical qubits and limited gate fidelity. An important question is how to solve some practical and meaningful tasks on such NISQ devices.

 Variational Quantum Algorithms (VQAs) are expected to realize quantum advantages on NISQ devices. VQAs are a class of hybrid quantum-classical algorithms. Specifically, VQAs employ a shallow-depth quantum circuit to efficiently evaluate a cost function which depends on the parameters of a quantum gate sequence on the quantum computer, and the classical computer uses this cost information to adjust the parameters of the gate sequence to minimize the cost function. VQAs have been successfully applied to calculate the ground state or the excited state of the system Hamiltonian \cite{PMS2014,KMT2017,JEM2019,HWB2019}, diagnose a quantum state \cite{LTO2019}, solve combinatorial optimization problems \cite{FGG2014,FH2016}, process classification tasks \cite{HCT2020}, solve linear systems \cite{XSE2019,HBR2019,BLC2019}, etc.

 Here, our aim is to design a VQA to solve the Poisson equation. A straight idea is to first adopt the finite difference method to discretize the Poisson equation to obtain a linear system, then employ the existing techniques \cite{XSE2019,HBR2019,BLC2019} directly to solve the linear system. However, the algorithms proposed in Refs.\cite{XSE2019,HBR2019,BLC2019} always need to satisfy the following conditions. Specifically speaking, $(1)$ the coefficient matrix of a linear system can be decomposed into a sum of tensor products, with $O(\polylog n)$ items, of a specific set of simple operators, where $n$ is the dimension of the coefficient matrix. And the smaller the number of decomposition items, the less quantum resources are required by the algorithm; $(2)$ the expectation values of each term of the tensor products of simple operators can be efficiently evaluated on a quantum computer. A common example is the decomposition of a coefficient matrix into a sum of tensor products of Pauli operators weighted by constant coefficients. Therefore, a key problem in designing a VQA for solve the linear system generated by the discrete Poisson equation is to find a decomposition that make the coefficient matrix $A$ satisfies the above requirements, but this decomposition is nontrivial. For example, the number of decomposed items of $A$ under the Pauli basis usually grows polynomially with the dimension of $A$.

 In this work, according to the special structure of the linear system, we find an explicit tensor product decomposition of its coefficient matrix $A$ under a set of simple operators $\{I,\sigma_{+}=|0\rangle\langle1|,\sigma_{-}=|1\rangle\langle0|\}$. It is worth emphasizing that the number of decomposition items is only $2\log n+1$, which means that the proposed VQA only needs fewer quantum measurements. Furthermore, we construct observables corresponding to the simple operators to efficiently evaluate the cost function on a quantum computer. The coefficient matrix $A$ satisfies the above two conditions, thus we design a VQA to solve the Poisson equation. Finally, we conduct numerical experiments to simulate our algorithm on ProjectQ \cite{SHT2018}, and the experimental results show that our algorithm can effectively solve the Poisson equation.

The remainder of the paper is organized as follows. In Sec.~\ref{Sec:Discretize}, we adopt the finite difference method to discretize the Poisson equation to obtain a linear system. In Sec.~\ref{Sec:Variational}, we propose a VQA for the Poisson equation and conduct numerical experiments to show the feasibility of our algorithm. Finally we present our conclusion in Sec.~\ref{Sec:conclusion}.

\section{ Discretize the Poisson equation}
\label{Sec:Discretize}
The $d$-dimensional Poisson equation with Dirichlet boundary conditions is defined as follows:
\begin{equation}
\begin{aligned}
\begin{split}
-\bigtriangleup \mu(x)&=f(x),&x\in D,\\
\mu(x)&=0,&x\in\partial D ,
\end{split}
\end{aligned}
\end{equation}
where $\bigtriangleup$ is the Laplace operator, $D:=(0,1)^{d}$ is the domain of $\mu(x)$, $\partial D $ represents the boundary of $D$ and $f: D\rightarrow R$ is a sufficiently smooth function \cite{ELC1998}. Here, we adopt the finite difference method to discretize the Poisson equation to obtain a linear system \cite{FGEW2008}, and then we obtain the approximate solution of the Poisson equation by solving the linear systems. The linear systems generated by the discretization of the 1-dimensional Poisson equation is:
\begin{equation}
A\mathbf{x}=\mathbf{b},
\end{equation}
where
\begin{equation}
A=\left[
\begin{array}{cccccc}
2 & -1 &    & 0 \\
-1 & \ddots & \ddots&  \\
  & \ddots& \ddots & -1 \\
0 &   & -1 & 2 \\
 \end{array}
 \right]\in R ^{n\times n},
 \end{equation}
 $n$ comes from dividing $(0,1)$ into $n+1$ parts evenly during discretization and $\mathbf{b}$ is the vector obtained by sampling the function $f(x)$ on the interior grid points \cite{DJW1997}.  Similarly, we can also obtain the coefficient matrix generated by the discretization of the d-dimensional Poisson equation as
 \begin{equation}
 \begin{aligned}
 \begin{split}
 A^{(d)}&=\underbrace{A\otimes I \otimes \cdots \otimes I}_{d}+I\otimes A\otimes I \otimes \cdots \otimes I+\cdots\\
 &+I \otimes \cdots \otimes I\otimes A,
 \end{split}
 \end{aligned}
 \end{equation}
where $I\in R ^{n\times n}$ and $A^{(d)}\in R^{n^{d}\times n^{d}}$.

\section{ A VQA for Poisson equation}
\label{Sec:Variational}
In order to design a VQA to solve the Poisson equation, we transform solving the linear systems that approximates the Poisson equation into finding the ground state of Hamiltonian $H$.
Here, $H$ can be constructed as:
 \begin{equation}
 H=A^{\dagger}(I-|\mathbf{b}\rangle\langle\mathbf{b}|)A=A(I-|\mathbf{b}\rangle\langle\mathbf{b}|)A,
 \end{equation}
 where the quantum state $|\mathbf{b}\rangle$ is proportional to the vector $\mathbf{b}$, which can be efficiently prepared by a unitary operator $U$ and the second equality comes from $A$ is a Hermitian matrix. It can be verified that $|\mathbf{x}\rangle=A^{-1}|\mathbf{b}\rangle/\|A^{-1}|\mathbf{b}\rangle\|$ is the ground state corresponding to its $0$ eigenvalue.

 To calculate the ground state $|\mathbf{x}\rangle$, we first find the ground state energy of $H$, which can be converted into find the minimum of the following cost function:
\begin{equation}
\begin{aligned}
\begin{split}
E(\bm{\theta})&=\langle\psi(\bm{\theta})|H|\psi(\bm{\theta})\rangle\\
&=\langle\psi(\bm{\theta})|A^{2}|\psi(\bm{\theta})\rangle-|\langle\mathbf{b}|A|\psi(\bm{\theta})\rangle|^{2},
\end{split}
 \end{aligned}
 \end{equation}
where $|\psi(\bm{\theta})\rangle$ is a parameterized quantum state, namely $|\psi(\bm{\theta})\rangle=U(\bm{\theta})|\mathbf{0}\rangle$, with unitary gate sequence $U(\bm{\theta})=U_{L}(\theta_{L})\cdots U_{1}(\theta_{1}), \bm{\theta}=(\theta_{L},\cdots,\theta_{1})$. When we obtain $\min_{\bm{\theta}} E(\bm{\theta})$ (that is, the ground state energy of $H$), and at the same time we obtain $\bm{\theta}=\bm{\theta}_{opt}$, then $U(\bm{\theta}_{opt})$ will produce the state $|\psi(\bm{\theta}_{opt})\rangle\approx|\mathbf{x}\rangle$. Therefore, the goal of our VQA is to obtain $\min_{\bm{\theta}} E(\bm{\theta})$. To this end, we need to evaluate $E(\bm{\theta})$ on the quantum computer, and adjust the parameters of the gate sequence $\bm{\theta}$ on the classical computer using the information of $E(\bm{\theta})$ to minimize $E(\bm{\theta})$.

\begin{figure}
\centering
\includegraphics[width=8.5cm]{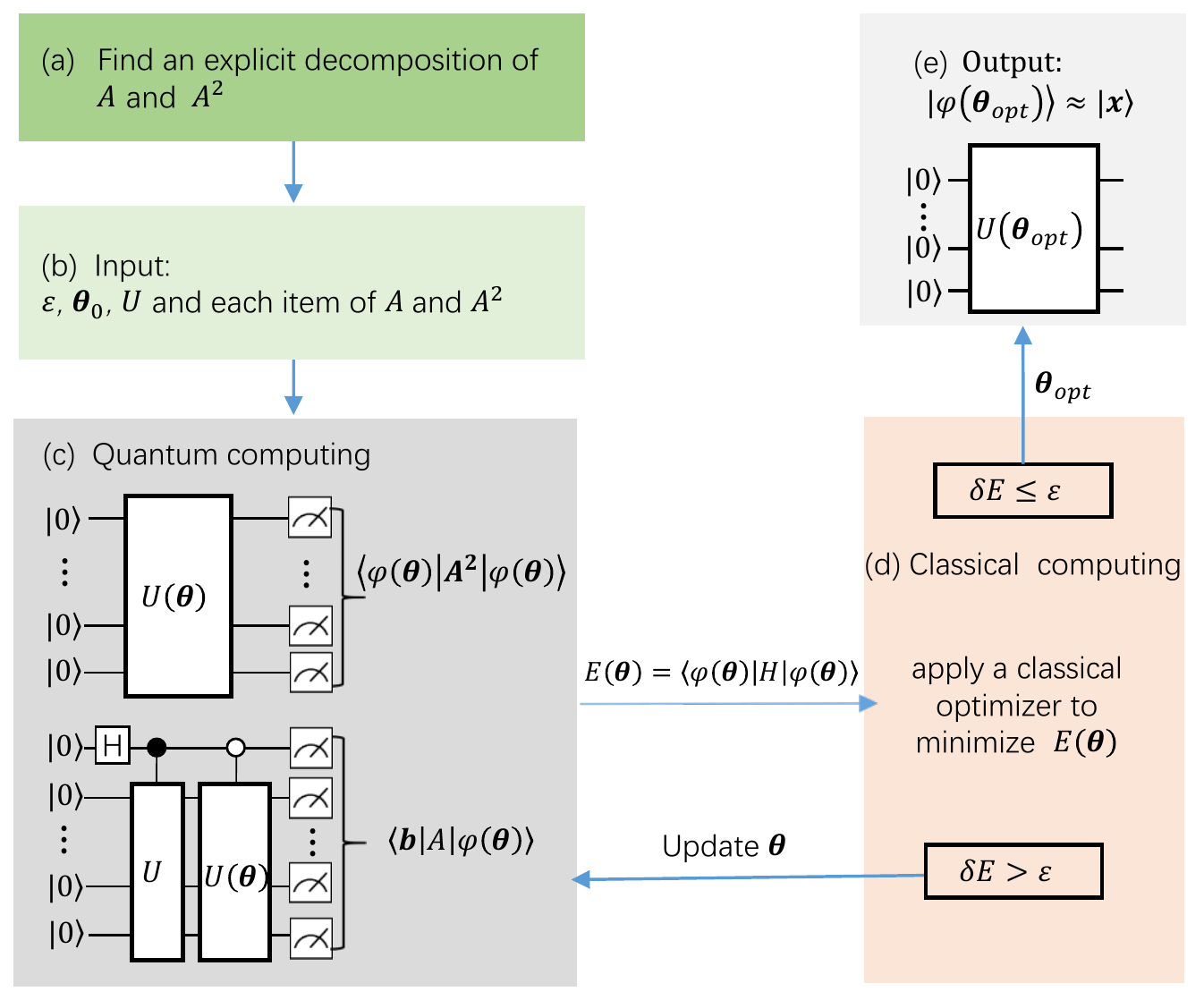}\\
\caption{Schematic diagram showing the steps of the entire algorithm. $(a)$ An explicit decomposition, which satisfies the two requirements mentioned in the introduction, of $A$ and $A^{2}$ under a specific set of simple operators are found. $(b)$ The inputs to our algorithm are the precision $\varepsilon$, the initial value $\bm{\theta}_{0}$, the unitary operator $U$ such that $U|\mathbf{0}\rangle=|\mathbf{b}\rangle$ and every item of a sum of tensor products of $A$ and $A^{2}$. $(c)$ To evaluate the cost function $E=E(\bm{\theta})$ on the quantum computer, we perform the unitary gate sequence $U(\bm{\theta})$ to evaluate $\langle\psi(\bm{\theta})|A^{2}|\psi(\bm{\theta})\rangle$ and perform $U$ and $U(\bm{\theta})$ to evaluate $\langle\mathbf{b}|A|\psi(\bm{\theta})\rangle$, respectively. $(d)$ We apply a classical optimizer (e.g., gradient descent) to minimize $E(\bm{\theta})$. If $\delta E>\varepsilon$, where $\delta E$ denote the change value of $E$, then update $\bm{\theta}$ to execute a new round of the quantum algorithm, otherwise output $\bm{\theta}_{opt}=\bm{\theta}$. $(e)$ The output of VQA is a quantum state $|\psi(\bm{\theta}_{opt}\rangle \approx$ $|\mathbf{x}\rangle$ generated by $U(\bm{\theta}_{opt})$.} \label{fig:process}
\end{figure}

To evaluate $E(\bm{\theta})$, we observe that $\langle\psi(\bm{\theta})|A^{2}|\psi(\bm{\theta})\rangle$ can be regard as the expectation value of observable $A^{2}$, and $|\langle\mathbf{b}|A|\psi(\bm{\theta})\rangle|^{2}$ can be seen as the square of the expectation value of observable $X\otimes A$ as follow:
\begin{equation}
\begin{aligned}
|\langle\mathbf{b},\psi(\bm{\theta})|X\otimes A|\mathbf{b},\psi(\bm{\theta})\rangle|^{2}=|{\langle\mathbf{b}|A|\psi(\bm{\theta}\rangle}|^{2},
\end{aligned}
\end{equation}
where $|\mathbf{b},\psi(\bm{\theta})\rangle=\frac{1}{\sqrt{2}}(|0\rangle|\mathbf{b}\rangle+|1\rangle|\psi(\bm{\theta})\rangle) $. When find a decomposition of $A$ and $A^{2}$ that satisfies the two requirements mentioned in the introduction, we can evaluate $E(\bm{\theta})$ on the quantum computer. The structure of the entire algorithm is shown in Fig.~\ref{fig:process}.

In the following sections, for convenience, we first consider the matrix $A$ generated by the discretization of the 1-dimensional Poisson equation and assume that $n=2^{m}$, where $m$ is a positive integer.


\subsection{An explicit decomposition of $A$ and $A^{2}$}
\label{Subsec:Decomposition}
We will show the process of obtaining a sum of the tensor products of $A$ and $A^{2}$ under a specific set of simple operators. According to the special structures of $A$ and $A^{2}$, we first write $A$ and $A^{2}$ into block matrices respectively, then apply the recursive algorithm to find their explicit decomposition.

Let's write $A$ as a block matrix:
 \begin{equation}
 A_{m}=\begin{bmatrix}
   \begin{array}{c | c}
   A_{m-1} & D_{m-1}   \\ \hline
  D^\mathrm{T}_{m-1} & A_{m-1}\\
   \end{array}
 \end{bmatrix},
 \end{equation}
 where
 \begin{equation}
  D_{m-1}=\left[
\begin{array}{cccc}
0  &  & & 0 \\
& \ddots & &   \\
0& & & \\
-1 & 0 & &0 \\
\end{array}
 \right].
 \end{equation}
Then we adopt the recursive algorithm to find the explicit decomposition of $A_{m}$ as follow:
\begin{equation}
\begin{aligned}
A_{1}&=\left[
\begin{array}{c|c}
2 & -1 \\ \hline
-1 & 2 \\
 \end{array}
 \right]=2I-\sigma_{+}-\sigma_{-};\\
A_{2}&=
\left[
\begin{array}{cc|cc}
2 & -1 & 0 & 0 \\
-1 & 2 & -1 & 0 \\ \hline
0 & -1 & 2 & -1 \\
0 & 0 & -1 & 2 \\
 \end{array}
 \right]\\
 &=I\otimes A_{1}-\sigma_{-}\otimes \sigma_{+}-\sigma_{+}\otimes \sigma_{-};\\
A_{3}&=I\otimes A_{2}-\sigma_{-}\otimes\sigma_{+}\otimes\sigma_{+}-\sigma_{+}\otimes
\sigma_{-}\otimes\sigma_{-}\\
&=I\otimes I\otimes(2I-\sigma_{+}-\sigma_{-})-I\otimes \sigma_{-}\otimes \sigma_{+}-I\otimes\sigma_{+}\otimes \sigma_{-}\\
&-\sigma_{-}\otimes\sigma_{+}\otimes\sigma_{+}-\sigma_{+}\otimes
\sigma_{-}\otimes\sigma_{-},
 \end{aligned}
 \end{equation}
 where $\sigma_{+}=|0\rangle\langle1|,\sigma_{-}=|1\rangle\langle0|$.
Then we have
\begin{widetext}
\begin{equation}
\begin{aligned}
\begin{split}
A_{m}&=I\otimes A_{m-1}-\sigma_{-}\otimes \underbrace{\sigma_{+}\otimes\cdots\otimes\sigma_{+}}_{m-1}-
\sigma_{+}\otimes\underbrace{\sigma_{-}\otimes\cdots\otimes\sigma_{-}}_{m-1}\\
&=\underbrace{I\otimes \cdots\otimes I}_{m-1}\otimes (2I-\sigma_{+}-\sigma_{-})-\underbrace{I\otimes \cdots\otimes I}_{m-2}\otimes \sigma_{-}\otimes \sigma_{+}-\cdots-I\otimes \sigma_{-}\otimes \underbrace{\sigma_{+}\otimes\cdots\otimes\sigma_{+}}_{m-1}\\
&-\underbrace{I\otimes \cdots\otimes I}_{m-2}\otimes \sigma_{+}\otimes \sigma_{-}-\cdots-I\otimes \sigma_{+}\otimes \underbrace{\sigma_{-}\otimes\cdots\otimes\sigma_{-}}_{m-2}-\sigma_{-}\otimes \underbrace{\sigma_{+}\otimes\cdots\otimes\sigma_{+}}_{m-1}-
\sigma_{+}\otimes\underbrace{\sigma_{-}\otimes\cdots\otimes\sigma_{-}}_{m-1}.
\end{split}
\end{aligned}
\end{equation}
\end{widetext}
It shows that $A_{m}$ can be written as a linear combination of tensor products of simple operators $\{I,\sigma_{+},\sigma_{-}\}$ and the total number of items of $A_{m}$ is $2m+1$, which is linear with respect to the logarithm of the dimension of the matrix. It means that our algorithm requires fewer quantum measurements, which will dramatically reduce quantum resources. Although the decomposition form of $A^{2}_{m}$ can be obtained by $A_{m}$, the number of terms is $(2m+1)^{2}$. In order to reduce the number of decomposition items, next we use the same method as $A_{m}$ to show the decomposition process of $A^{2}_{m}$.

 $A^{2}_{m}$ is shown as follows:
\begin{equation}
\begin{aligned}
\begin{split}
 A^{2}_{m}&=\left[
\begin{array}{cccccc}
5 & -4 & 1& &  &0 \\
-4 & 6 & -4 & 1&  & \\
1 & \ddots& \ddots & \ddots& \ddots & \\
 & \ddots & \ddots &\ddots & \ddots & 1\\
 &  & 1&-4 & 6 & -4 \\
0&   &  &1 & -4 & 5 \\
 \end{array}
 \right]\\
 = &\left[
\begin{array}{cccccc}
6 & -4 & 1& &  &0 \\
-4 & 6 & -4 & 1 &  & \\
1 & \ddots& \ddots & \ddots& \ddots & \\
 & \ddots & \ddots &\ddots & \ddots & 1\\
 &  & 1&-4 & 6 & -4 \\
0&   &  &1 & -4 & 6 \\
 \end{array}
 \right]-
  \left[
\begin{array}{cccccc}
1&  & & & & 0 \\
 & 0& & &  & \\
 &  & \ddots & & & \\
 &  & & 0 & &  \\
 &  & & & 0 & \\
0&  & & & & 1\\
 \end{array}
 \right]\\
 &\equiv B_{m}-C_{m}.
 \end{split}
 \end{aligned}
 \end{equation}
According to Eq.$(12)$, we only need to obtain the decomposition of $B_{m}$ and $C_{m}$. We write $B_{m}$ into a block matrix :
\begin{equation}
 B_{m}=\begin{bmatrix}
   \begin{array}{c | c}
    B_{m-1}& M_{m-1}  \\ \hline
  M^\mathrm{T}_{m-1} & B_{m-1}\\
   \end{array}
 \end{bmatrix},
 \end{equation}
where
\begin{equation}
M_{m-1}= \left[
\begin{array}{ccccc}
 0& & &  &0\\
 & & &  &\\
0& &\ddots & & \\
1&0 & & & \\
-4 & 1 &0& &0\\
\end{array}
 \right].
 \end{equation}
 Next, we apply the recursive algorithm to obtain the decomposition of $B_{m}$:

 \begin{equation}
\begin{aligned}
\begin{split}
B_{1}&=\left[
\begin{array}{c|c}
6 & -4 \\ \hline
-4& 6 \\
 \end{array}
 \right]=6I-4\sigma_{+}-4\sigma_{-};\\
B_{2}&=
\left[
\begin{array}{cc|cc}
6 & -4 & 1 & 0 \\
-4 & 6 & -4 & 1\\ \hline
1 & -4 & 6 & -4 \\
0 & 1 & -4 & 6\\
 \end{array}
 \right]\\
&=I\otimes B_{1}+\sigma_{-}\otimes (I-4\sigma_{+})+\sigma_{+}\otimes (I-4\sigma_{-});\\
 B_{3}&=I\otimes B_{2}+\sigma_{-}\otimes \sigma_{+}\otimes(I-4\sigma_{+})+\sigma_{+}\otimes \sigma_{-}\otimes(I-4\sigma_{-})\\
&=I\otimes I\otimes(6I-4\sigma_{+}-4\sigma_{-})\\
&+I\otimes \sigma_{-}\otimes (I-4\sigma_{+})+I\otimes \sigma_{+}\otimes (I-4 \sigma_{-})\\
&+\sigma_{-}\otimes \sigma_{+}\otimes(I-4\sigma_{+})+\sigma_{+}\otimes \sigma_{-}\otimes(I-4\sigma_{-}).\\
\end{split}
 \end{aligned}
 \end{equation}
And then we can obtain
\begin{widetext}
\begin{equation}
\begin{aligned}
\begin{split}
B_{m}&=I\otimes B_{m-1}+\sigma_{-}\otimes \underbrace{\sigma_{+}\otimes\cdots\otimes\sigma_{+}}_{m-2}\otimes(I-4\sigma_{+})+\sigma_{+}\otimes \underbrace{\sigma_{-}\otimes\cdots\otimes\sigma_{-}}_{m-2}\otimes(I-4\sigma_{-})\\
&=\underbrace{I\otimes \cdots\otimes I}_{m-1}\otimes(6I-4\sigma_{+}-4\sigma_{-})+\underbrace{I\otimes \cdots\otimes I}_{m-2}\otimes \sigma_{+}\otimes(I-4\sigma_{-})+\cdots+I\otimes\sigma_{+}\otimes \underbrace{\sigma_{-}\otimes\cdots\otimes\sigma_{-}}_{m-3}\otimes(I-4\sigma_{-})\\
&+\underbrace{I\otimes \cdots\otimes I}_{m-2}\otimes \sigma_{-}\otimes(I-4\sigma_{+})+\cdots+I\otimes\sigma_{-}\otimes \underbrace{\sigma_{+}\otimes\cdots\otimes\sigma_{+}}_{m-3}\otimes(I-4\sigma_{+})\\
&+\sigma_{-}\otimes \underbrace{\sigma_{+}\otimes\cdots\otimes\sigma_{+}}_{m-2}\otimes(I-4\sigma_{+})+\sigma_{+}\otimes \underbrace{\sigma_{-}\otimes\cdots\otimes\sigma_{-}}_{m-2}\otimes(I-4\sigma_{-}).\\
 \end{split}
 \end{aligned}
 \end{equation}
 \end{widetext}
 Thus $B_{m}$ can be expressed as a sum of tensor products of operators $\{I,\sigma_{+},\sigma_{-}\}$ and the number of items is $4m-1$. Finally, we obtain the decomposition form of $C_{m}$ via the recursive algorithm:
\begin{equation}
\begin{aligned}
\begin{split}
C_{1}&=\left[
\begin{array}{c|c}
1 & 0 \\ \hline
0 & 1\\
 \end{array}
 \right]=\sigma_{+} \sigma_{-}+\sigma_{-} \sigma_{+};\\
C_{2}&=
\left[
\begin{array}{cc|cc}
1 & 0 & 0 & 0 \\
0 & 0 & 0 & 0\\ \hline
0 & 0 & 0 & 0 \\
0 & 0 & 0 & 1\\
 \end{array}
 \right]=\sigma_{+} \sigma_{-}\otimes\sigma_{+} \sigma_{-}+\sigma_{-} \sigma_{+}\otimes\sigma_{-} \sigma_{+};\\
C_{3}&=\sigma_{+}\sigma_{-}\otimes\sigma_{+}\sigma_{-}\otimes\sigma_{+} \sigma_{-}+\sigma_{-} \sigma_{+}\otimes\sigma_{-} \sigma_{+}\otimes\sigma_{-} \sigma_{+}.\\
\end{split}
 \end{aligned}
 \end{equation}
 And we can obtain
\begin{equation}
C_{m}=\underbrace{\sigma_{+}\sigma_{-}\otimes\cdots\otimes\sigma_{+} \sigma_{-}}_{m}+\underbrace{\sigma_{-}\sigma_{+}\otimes\cdots\otimes\sigma_{-}\sigma_{+}}_{m}.
\end{equation}
It shows that $C_{m}$ is presented in the form of a sum of the tensor product of $\{\sigma_{+}\sigma_{-}, \sigma_{-}\sigma_{+}\}$. Thus the explicit decomposition form of $A^{2}_{m}$ can be obtained and the total number of items is $4m+1$. Similarly, we can also obtain the decomposition of $A^{(d)}$, with only $d(2m+1)$ items, generated by the discretization of the $d$-dimensional Poisson equation. Besides, we apply the decomposition method to the general tridiagonal and pentadiagonal Toeplitz matrices which are often used to solve partial differential equations \cite{ELC1998,C2007,FGEW2008,DJW1997} in Appendix.~\ref{Sec:Apply}.

\subsection{Evaluate $E(\bm{\theta})$}
\label{Subsec:Evaluate}
 In order to effectively evaluate $E(\bm{\theta})$, we directly perform quantum measurements to evaluate the expectation value of each item of $X\otimes A$ and $A^{2}=B-C$.

\begin{figure}
\centering
\includegraphics[width=8cm]{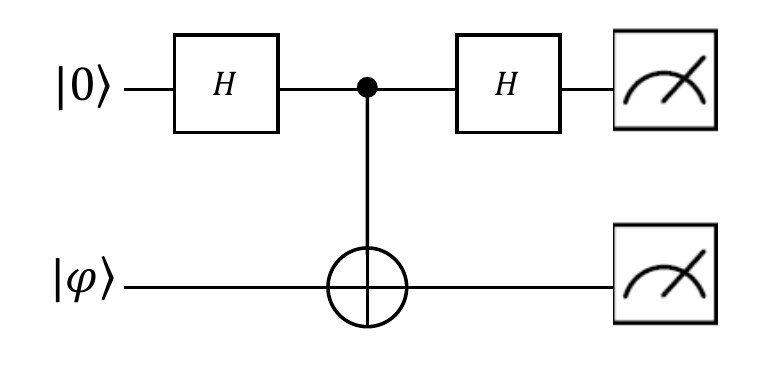}\\
\caption{A quantum circuit of the Bell measurement on a quantum computer. The top wire represents an ancilla qubit and the lower wire represents an arbitrary single-qubit state $|\varphi\rangle$.}\label{fig:bell}
\end{figure}

 According to the explicit decomposition of $A$ and $B$, we need to construct observables to evaluate their expected value. Here, observables can be designed as follows:
 \begin{equation}
 \begin{aligned}
 \begin{split}
  P_{+}&= \left[
\begin{array}{cc}
\mathbf{0} & \sigma_{+} \\
\sigma^{\dagger}_{+} & \mathbf{0} \\
 \end{array}
 \right]=|\phi^{+}\rangle\langle\phi^{+}|-|\phi^{-}\rangle\langle\phi^{-}|,\\
 P_{-}&= \left[
\begin{array}{cc}
\mathbf{0} & \sigma_{-} \\
\sigma^{\dagger}_{-} & \mathbf{0} \\
 \end{array}
 \right]=|\psi^{+}\rangle\langle\psi^{+}|-|\psi^{-}\rangle\langle\psi^{-}|,
 \end{split}
 \end{aligned}
 \end{equation}
 where $|\phi^{\pm}\rangle=\frac{1}{\sqrt{2}}(|00\rangle\pm|11\rangle)$ and $|\psi^{\pm}\rangle=\frac{1}{\sqrt{2}}(|01\rangle\pm|10\rangle)$ are Bell states. Then we can attach an qubit to directly perform quantum Bell measurements to obtain:
 \begin{equation}
 \langle+|\langle\varphi|P_{\pm}|+\rangle|\varphi\rangle=\langle\varphi|\sigma_{\pm}|\varphi\rangle,
 \end{equation}
 where $|+\rangle=\frac{1}{\sqrt{2}}(|0\rangle+|1\rangle)$ and $|\varphi\rangle$ is an arbitrary single-qubit state. Thus the expectation values of $X\otimes A$ and $B$ can be calculated from the result of the Bell measurements which can be done in parallel for all pairs of qubits using Hardmard and CNOT gates \cite{JKMN2020}. The quantum circuit is shown in Fig.~\ref{fig:bell}.

To evaluate the expectation value of each item of $C$, we note that $\sigma_{-}\sigma_{+}=|1\rangle\langle1|$ and $\sigma_{+}\sigma_{-}=|0\rangle\langle0|$ are Hermitian operators. Thus, we can perform quantum local measurements directly on the computational basis to obtain the expectation value of $C$. Coupled with the linearity property of operators, we can evaluate the cost function $E(\bm{\theta})$ efficiently on quantum computer.

\subsection{ Numberical experiments}
\label{subsec:numberical}
\begin{figure}
\centering
\includegraphics[width=8cm]{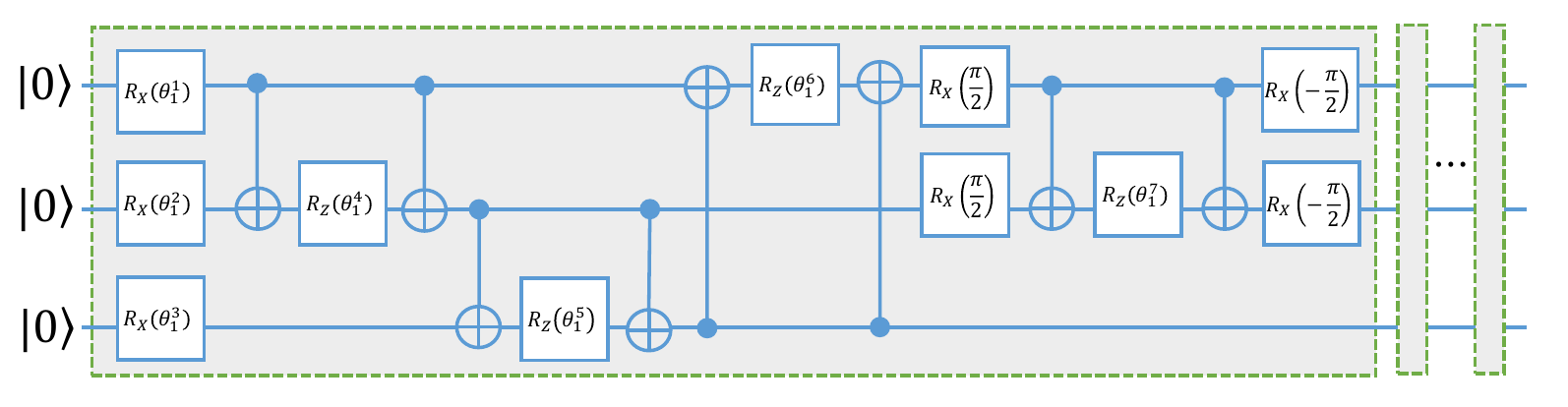}\\
\caption{A variational quantum circuit with $m=3$ qubit. Here $R_{X}$ and $R_{Z}$ are single qubit rotations around the $X$ and $Z$ axes, respectively. $\theta_{i}^{j}$ represents the $j$th parameter of the $i$th layer, $i=1,\cdots,8,j=1,\cdots,7$. The light green dashed box indicates the repeated block.}\label{fig:n=3qubit}
\end{figure}

\begin{figure}
\centering
\includegraphics[width=8cm]{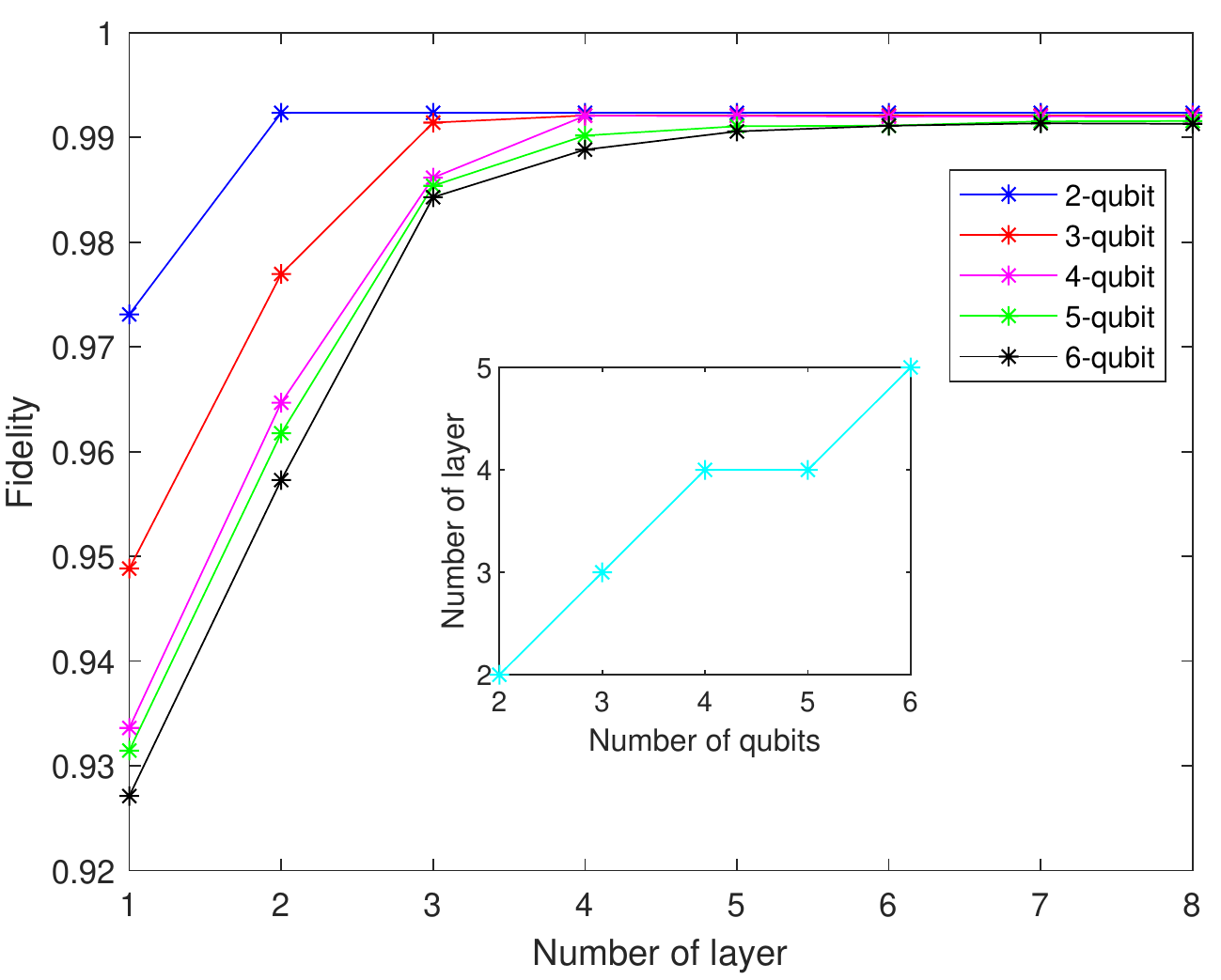}
\caption{The fidelity $|\langle\mathbf{x}|\psi(\bm{\theta}_{opt})\rangle|$ with increasing number of layers and number of qubits. For a given number
of qubits, the layer of quantum circuit is increased gradually until the fidelity reaches $0.99$. The inserted graph shows the minimum layer of quantum circuit in the simulation when the fidelity reaches $0.99$. }\label{fig:fidelity}
\end{figure}

We conduct numerical experiments to simulate our algorithm in ProjectQ \cite{SHT2018}, which is a high-performance simulator with emulation capabilities. In our experiments, we consider the system Hamiltonian with $m=2,\cdots,6$ qubits corresponding to the 1-dimensional Poisson equation, where the vector $\mathbf{b}$ we choice is obtained by sampling the function $f(x)=x$ on the interior grid points. It is worth noting that the solution of the linear system get closer to the analytic solution of the Poisson equation as the dimension of the linear system increase \cite{ELC1998,C2007,FGEW2008,DJW1997}. Next we design a variational circuit $U(\bm{\theta})$ to generate the quantum state $|\psi(\bm{\theta})\rangle$. Here, we apply the Quantum Alternating Operator Ansatz (QAOA) \cite{FGG2014,SZBEDR2019} to design $U(\bm{\theta})$. The QAOA consists of evolving the $|+\rangle^{\otimes n}$ state by a driver Hamiltonian $H_{D}$ and a mixer Hamiltonian $H_{M}$ for a specified number of layers $p$. The variational circuit $U(\bm{\theta})$ is obtain by alternating the unitary operators $U_{D}(\theta_{i}^{j}):= \exp(-iH_{D}\theta_{i}^{j})$ and $U_{M}(\theta_{i+1}^{j}):= \exp(-iH_{M}\theta_{i+1}^{j})$ $p$ times:
\begin{equation}
U(\bm{\theta})=U_{M}(\theta_{L}^{p})U_{D}(\theta_{L-1}^{p})\cdots U_{M}(\theta_{2}^{1}) U_{D}(\theta_{1}^{1}),
\end{equation}
where $\bm{\theta}=(\theta_{L}^{p},\theta_{L-1}^{p},\cdots, \theta_{2}^{1}, \theta_{1}^{1})$, $\theta_{i}^{j}$ represents the $j$th parameter of the $i$th layer, $i=1,\cdots,L,j=1,\cdots,p$. In our numerical experiments, $H_{M}=\sum_{i=1}^{n}X_{i}$ and $H_{D}=\sum_{i=0}^{n}Z_{i}Z_{i+1}+Z_{n}Z_{0}+Y_{0}Y_{1}$, where $X$,$Y$ and $Z$ are Pauli operators. And each $\theta_{i}^{j}\in [0,2\pi)$ of $\bm{\theta}$ is chosen randomly. We adopt the Broyden-Fletcher-Goldfarb-Shanno (BFGS) as the optimizer. In Fig.~\ref{fig:n=3qubit}, we give an example of the quantum circuit with $m=3$ qubit. From Fig.~\ref{fig:fidelity}, we observe that as the layers of quantum circuit $p$ increases, the fidelity $|\langle\mathbf{x}|\psi(\bm{\theta}_{opt})\rangle|$ of quantum circuits gradually increases, and can reach $0.99$. We also obtain the minimum layers of quantum circuits that is required to guarantee the fidelity $0.99$, plotted as an insert.

\section{conclusion}
\label{Sec:conclusion}
 To summarize, we designed a VQA to solve the Poisson equation. In particular, we found an explicit decomposition of the coefficient matrix of the linear system that approximates the Poisson equation. It is noteworthy that the number of decomposition items is only $2\log n+1$, where $n$ is the dimension of the coefficient matrix, which greatly reduces the number of measurements in VQAs. In addition, we performed quantum Bell measurements in parallel to evaluate the cost function on a quantum computer.

 Besides, we applied the decomposition method to the general tridiagonal and pentadiagonal Toeplitz matrices which are often utilized in solving partial differential equations \cite{ELC1998,C2007,FGEW2008,DJW1997}. And the number of decomposition items grows linearly with the logarithm of the dimension of the matrix. The banded Toeplitz systems with bandwidth $p$ have wide applications in many fields, such as solving the partial differential equations \cite{ELC1998,C2007,FGEW2008,DJW1997}, Markov chains \cite{M1981,M1989} and signal processing \cite{JR1996}, where $p\geq2$ is a positive integer. Similarly, our algorithm also can be extended to address the banded Toeplitz systems.

 When designing VQAs to solve some practical problems, such as dimensionality reduction, classification and linear system, the data matrix usually needs to meet the two requirements mentioned in the introduction. In the past, we often choose to decompose the data matrix under the Pauli operators. Our algorithm provides a new decomposition idea that the data matrix can be decomposed into a sum of tensor products of a class of simple operators which may not be Pauli operators or even Hermitian operators, as long as the expectation value of such operators can be measured efficiently on a quantum computer. Our algorithm may stimulate more VQAs for solving practical application problems.

\begin{acknowledgments}
This work is supported by NSFC (Grants No.61976024, No. 61972048) and the Fundamental Research Funds for the Central Universities (Grant No.2019XD-A01).\\
\end{acknowledgments}

 \begin{widetext}
\appendix
\section{Decomposition of the general tridiagonal and pentadiagonal Toeplitz matrices}
\label{Sec:Apply}
 In this Appendix, we apply the decomposition method of our algorithm to the general tridiagonal and pentadiagonal Toeplitz matrices, which often appear in solving partial differential equations \cite{ELC1998,C2007,FGEW2008,DJW1997}. Here we assume that $n=2^{m}$, where $m$ is a positive integer. The general Tridiagonal and pentadiagonal Toeplitz matrices are defined as follows:
 \begin{equation}
W=\left[
\begin{array}{cccccc}
t_{0} & t_{1} &    & 0 \\
t_{-1} & \ddots & \ddots&  \\
  & \ddots& \ddots & t_{1} \\
0 &   & t_{-1} & t_{0}\\
 \end{array}
 \right]\in R ^{n\times n},\\
 V= \left[
\begin{array}{cccccc}
t_{0}& t_{1}& t_{2}& &  &0 \\
t_{-1}&t_{0} &t_{1} & t_{2}&  & \\
t_{-2} & \ddots& \ddots & \ddots& \ddots & \\
 & \ddots & \ddots &\ddots & \ddots & t_{2}\\
 &  &t_{-2}& t_{-1}&t_{0} &t_{1} \\
0&   &  &t_{-2} & t_{-1} & t_{0} \\
 \end{array}
 \right]\in R ^{n\times n},
 \end{equation}
 Then we adopt the recursive algorithm to find an explicit tensor product decomposition of $W$ and $V$ under a set of simple operators $\{I,\sigma_{+},\sigma_{-}\}$ respectively, which can be expressed as:
\begin{equation}
\begin{aligned}
\begin{split}
W&=\underbrace{I\otimes \cdots\otimes I}_{m-1}\otimes(t_{0}I+t_{-1}\sigma_{+}+t_{1}\sigma_{-})\\
&+\underbrace{I\otimes \cdots\otimes I}_{m-2}\otimes \sigma_{-}\otimes(t_{1}\sigma_{+})+\cdots+I\otimes \sigma_{-}\otimes\underbrace{ \sigma_{+}\cdots\otimes\sigma_{+}}_{m-3}\otimes(t_{1}\sigma_{+})+\sigma_{-}\otimes\underbrace{\sigma_{+}\otimes\cdots\otimes\sigma_{+}}_{m-2}\otimes(t_{1}\sigma_{+})\\
&+\underbrace{I\otimes \cdots\otimes I}_{m-2}\otimes \sigma_{+}\otimes (t_{-1}\sigma_{-})+\cdots+I\otimes \sigma_{+}\otimes\underbrace{\sigma_{-}\cdots\otimes\sigma_{-}}_{m-3}\otimes(t_{-1}\sigma_{-})+\sigma_{+}\otimes\underbrace{\sigma_{-}\otimes\cdots\otimes\sigma_{-}}_{m-2}\otimes(t_{-1}\sigma_{-}),\\
V&=\underbrace{I\otimes \cdots\otimes I}_{m-1}\otimes (t_{0}I+t_{-1}\sigma_{+}+t_{1}\sigma_{-})\\
&+\underbrace{I\otimes \cdots\otimes I}_{m-2}\otimes \sigma_{-}\otimes (t_{2}I+t_{1}\sigma_{+})+\cdots+I\otimes \sigma_{-}\otimes \underbrace{\sigma_{+}\otimes\cdots\otimes\sigma_{+}}_{m-3}\otimes (t_{2}I+t_{1}\sigma_{+})+\sigma_{-}\otimes\underbrace{\sigma_{+}\otimes\cdots\otimes\sigma_{+}}_{m-2}\otimes (t_{2}I+t_{1}\sigma_{+})\\
&+\underbrace{I\otimes \cdots\otimes I}_{m-2}\otimes \sigma_{+}\otimes (t_{-2}I+t_{-1}\sigma_{-})+\cdots+I\otimes \sigma_{+}\otimes \underbrace{\sigma_{-}\otimes\cdots\otimes\sigma_{-}}_{m-3}\otimes (t_{-2}I+t_{-1}\sigma_{-})\\
&+\sigma_{+}\otimes\underbrace{\sigma_{-}\otimes\cdots\otimes\sigma_{-}}_{m-2}\otimes (t_{-2}I+t_{-1}\sigma_{-}).\\
 \end{split}
 \end{aligned}
 \end{equation}
 And the total number of decomposed items of $W$ and $V$ are $2m+1$ and $4m-1$, respectively.
  \end{widetext}

\nocite{*}
\bibliography{apssamp}

\begin{thebibliography}{0}%
\makeatletter
\providecommand \@ifxundefined [1]{%
 \@ifx{#1\undefined}
}%
\providecommand \@ifnum [1]{%
 \ifnum #1\expandafter \@firstoftwo
 \else \expandafter \@secondoftwo
 \fi
}%
\providecommand \@ifx [1]{%
 \ifx #1\expandafter \@firstoftwo
 \else \expandafter \@secondoftwo
 \fi
}%
\providecommand \natexlab [1]{#1}%
\providecommand \enquote  [1]{``#1''}%
\providecommand \bibnamefont  [1]{#1}%
\providecommand \bibfnamefont [1]{#1}%
\providecommand \citenamefont [1]{#1}%
\providecommand \href@noop [0]{\@secondoftwo}%
\providecommand \href [0]{\begingroup \@sanitize@url \@href}%
\providecommand \@href[1]{\@@startlink{#1}\@@href}%
\providecommand \@@href[1]{\endgroup#1\@@endlink}%
\providecommand \@sanitize@url [0]{\catcode `\\12\catcode `\$12\catcode
  `\&12\catcode `\#12\catcode `\^12\catcode `\_12\catcode `\%12\relax}%
\providecommand \@@startlink[1]{}%
\providecommand \@@endlink[0]{}%
\providecommand \url  [0]{\begingroup\@sanitize@url \@url }%
\providecommand \@url [1]{\endgroup\@href {#1}{\urlprefix }}%
\providecommand \urlprefix  [0]{URL }%
\providecommand \Eprint [0]{\href }%
\providecommand \doibase [0]{http://dx.doi.org/}%
\providecommand \selectlanguage [0]{\@gobble}%
\providecommand \bibinfo  [0]{\@secondoftwo}%
\providecommand \bibfield  [0]{\@secondoftwo}%
\providecommand \translation [1]{[#1]}%
\providecommand \BibitemOpen [0]{}%
\providecommand \bibitemStop [0]{}%
\providecommand \bibitemNoStop [0]{.\EOS\space}%
\providecommand \EOS [0]{\spacefactor3000\relax}%
\providecommand \BibitemShut  [1]{\csname bibitem#1\endcsname}%
\let\auto@bib@innerbib\@empty
\end{thebibliography}%


\begin{thebibliography}{}\label{sec:TeXbooks}
\bibitem{S1994}P. W. Shor. Algorithms for quantum computation: Discrete logarithms and factoring, in \emph{Proceedings of the 35th Annual Symposium on the Foundations of Computer Science}, edited by S. Goldwasser (IEEE, Los Alamitos, CA), pp.124-134 (1994).
\bibitem{LKG1997}L. K. Grover. Quantum Mechanics Helps in Searching for a Needle in a Haystack. Phys. Rev. Lett. 79, 325 (1997).
\bibitem{HHL2009}A. W. Harrow, A. Hassidim, and S. Lloyd. Quantum algorithm for linear systems of equations. Phys. Rev. Lett. 103, 150502 (2019).
\bibitem{WYPGWQ2018}L. C. Wan, C. H. Yu, S. J. Pan, F. Gao, Q. Y. Wen, and S. J. Qin. Asymptotic quantum algorithm for the Toeplitz systems. Phys. Rev. A. 97, 062322 (2018).
\bibitem{PMSL2014}P. Rebentrost, M. Mohseni, and S. Lloyd. Quantum support vector machine for big data classification. Phys. Rev. Lett. 113, 130503 (2014).
\bibitem{BJYD2017}B. J. Duan, J. B. Yuan, Y. Liu, and D. Li. Quantum algorithm for support matrix machines. Phys. Rev. A. 96, 032301 (2017).
\bibitem{NDS2012}N. Wiebe, D. Braun, and S. Lloyd. Quantum algorithm for data fitting. Phys. Rev. Lett. 109, 050505 (2012).
\bibitem{YGW2019}C. H. Yu, F. Gao , and Q. Y. Wen. An improved quantum algorithm for ridge regression. IEEE Transactions on Knowledge and Data Engineering (2019).
\bibitem{SMP2014}S. Lloyd, M. Mohseni, and P. Rebentrost. Quantum principal component analysis. Nat. Phys. 10, 631 (2014).
\bibitem{IL2016}I. Cong and L. Duan. Quantum discriminant analysis for dimensionality reduction and classification. New J. Phys. 18, 073011 (2016).
\bibitem{PWL2020}S. J. Pan, L. C. Wan, H. L. Liu, Q. L. Wang, S. J. Qin, Q. Y. Wen, and F. Gao. An improved quantum algorithm for A-optimal projection. Phys. Rev. A. 102, 052402 (2020).
\bibitem{P2018}J. Preskill. Quantum Computing in the NISQ era and beyond. Quantum, 2, 79 (2018).
\bibitem{PMS2014}A. Peruzzo, J. McClean, P. Shadbolt, et al. A variational eigenvalue solver on a photonic quantum processor. Nat. Commun. 5, 4213 (2014).
\bibitem{KMT2017}A. Kandala, A. Mezzacapo, K. Temme, et al. Hardware-efficient variational quantum eigensolver for small molecules and quantum magnets. Nature. 549, 7671 (2017).
\bibitem{JEM2019}T. Jones, S. Endo, S. McArdle, et al. Variational quantum algorithms for discovering Hamiltonian spectra. Phys. Rev. A. 99, 062304 (2019).
\bibitem{HWB2019}O. Higgott, D. Wang, and S. Brierley. Variational quantum computation of excited states. Quantum. 3, 156 (2019).
\bibitem{LTO2019}R. LaRose, A. Tikku, \'{E}. O'Neel-Judy, et al. Variational quantum state diagonalization. npj Quantum Information. 5, 1-10 (2019).
\bibitem{FGG2014}E. Farhi, J. Goldstone, and S. Gutmann. A quantum approximate optimization algorithm. arXiv preprint arXiv: 1411. 4028 (2014).
\bibitem{FH2016}E. Farhi, and A. W. Harrow. Quantum supremacy through the quantum approximate optimization algorithm. arXiv preprint arXiv: 1602. 07674 (2016).
\bibitem{HCT2020}V. Havl\'{i}\v{c}ek , A. D. C\'{o}rcoles, K. Temme, et al. Supervised learning with quantum-enhanced feature spaces. Nature. 567, (7747) (2019).
\bibitem{TMC2005}J. Tomasi, B. Mennucci,  and R. Cammi. Quantum mechanical continuum solvation models. Chem. Rev. 105, 2999-3094 (2005).
\bibitem{MSP2007}S. P. Meyn. Control Techniques for Complex Networks. (Cambridge: Cambridge University Press) (2007).
\bibitem{AG2007}S. Asmussen, and P. W. Glynn. Stochastic Simulation: Algorithms and Analysis (Stochastic Modelling and Applied Probability vol 57) (Berlin: Springer) (2007).
\bibitem{FGEW2008}G. E. Forsythe, and W. R. Wasow. \emph{Finite-Difference Methods for Partial Differential Equations}. (New York: Dover) (2004).
\bibitem{C2007}C. I. Gheorghiu. \emph{Spectral Methods for Differential Problems}, Casa C\v{a}rtii de Stiint\v{a}, Cluj-Napoca, Romania (2007).
\bibitem{ELC1998}L. C. Evans. \emph{Partial Differential Equations} (Providence, RI: American Mathematical Society) (1998).
\bibitem{DJW1997}J. W. Demmel. \emph{Applied Numerical Linear Algebra} (Philadelphia, PA: SIAM) (1997).
\bibitem{CPP2019}Y. Cao, A. Papageorgiou, I. Petras, et al. Quantum algorithm and circuit design solving the Poisson equation. New J. Phys. 15, 013021 (2013).
\bibitem{CLO2020}A. M. Childs, J. P. Liu, and A. Ostrander. High-precision quantum algorithms for partial differential equations. arXiv preprint arXiv: 2002. 07868 (2020).
\bibitem{XSE2019}A. Xu, J. Sun, S. Endo S, et al. Variational algorithms for linear algebra. arXiv preprint arXiv: 1909. 03898 (2019).
\bibitem{HBR2019}H. Y. Huang, K. Bharti, and P.Rebentrost. Near-term quantum algorithms for linear systems of equations. arXiv preprint arXiv: 1909. 07344 (2019).
\bibitem{BLC2019}C. Bravo-Prieto, R. LaRose, M. Cerezo, et al. Variational Quantum Linear Solver: A Hybrid Algorithm for Linear Systems. arXiv preprint arXiv: 1909. 05820 (2019).
\bibitem{SSO2019}Y. Suba\c{s}{\i}, R. D. Somma, and D. Orsucci. Quantum algorithms for systems of linear equations inspired by adiabatic quantum computing. Phys. Rev. Lett. 122, 060504 (2019).
\bibitem{SHT2018}D. S. Steiger, T. H\"{a}ner, and M. Troyer. ProjectQ: an open source software framework for quantum computing. Quantum. 2, 49 (2018).
\bibitem{SZBEDR2019}S. Hadfield, Z. Wang, B. O'Gorman, E. G. Rieffel, D. Venturelli, and R. Biswas. From the quantum approximate optimization algorithm to a quantum alternating operator ansatz. \emph{Algorithms}. 12, 34 (2019).
\bibitem{M1981}M. F. Neuts. Matrix-Geometric Solutions in Stochastic Models. Johns Hopkins University Press, Baltimore (1981).
\bibitem{M1989}M. F. Neuts. Structured Stochastic Matrices of M/G/1 Type and Their Applications, Marcel Dekker, New York (1989).
\bibitem{JR1996}R. H. Chan, and M. K. Ng, Scientific applications of iterative Toeplitz solvers, Calcolo, 33, pp. 249-267 (1996).
\bibitem{JKMN2020}Z. Jiang, A. Kalev, W. Mruczkiewicz, and H. Neven. Optimal fermion-to-qubit mapping via ternary trees with applications to reduced quantum states learning. Quantum. 4, 276 (2020).
\end{thebibliography}
\end{document}